**EarthFinder: A Precise Radial Velocity Probe Mission Concept For the Detection of Earth-Mass Planets Orbiting Sun-like Stars**

Lead Author: Peter Plavchan, George Mason University, 626-234-1628, pplavcha@gmu.edu

Co-Authors:
Bryson Cale, Patrick Newman, Bahaa Hamze, Natasha Latouf, William Matzko
                                                                George Mason University

| | |
|---|---|
| Chas Beichman, David Ciardi | NASA Exoplanet Science Institute |
| Bill Purcell, Paul Lightsey | Ball Aerospace |
| Heather Cegla, Xavier Dumusque, Vincent Bourrier | University of Geneva |
| Courtney Dressing, Peter Gao | University of California, Berkeley |
| Gautam Vasisht, Stephanie Leifer | Jet Propulsion Laboratory |
| Sharon Wang, Jonathan Gagne | Carnegie DTM |
| Samantha Thompson | University of Cambridge Cambridge |
| Jonathan Crass, Andrew Bechter, Eric Bechter | Notre Dame |
| Cullen Blake, Sam Halverson | University of Penn |
| Andrew Mayo | Technical University of Denmark |
| Thomas Beatty, Jason Wright | Penn State |
| Alex Wise | University of Delaware |
| Angelle Tanner | Mississippi State |
| Jason Eastman, Sam Quinn | Harvard |
| Debra Fischer, Sarbani Basu, Sophia Sanchez-Maes | Yale |
| Andrew Howard, Kerry Vahala, Ji Wang | Caltech |
| Scott Diddams, Scott Papp | NIST |
| Benjamin JS Pope | New York University |
| Emily Martin | UCLA |
| Simon Murphy | University of Sydney |

**Summary:**
EarthFinder is a Probe Mission concept selected for study by NASA for input to the 2020 astronomy decadal survey. This study is currently active and a final white paper report is due to NASA at the end of calendar 2018.

We are tasked with evaluating the *scientific rationale* for obtaining precise radial velocity (PRV) measurements in space, which is a two-part inquiry:
- What can be gained from going to space?
- What can't be done from the ground?

These two questions flow down to these specific tasks for our study:
- Identify the velocity limit, if any, introduced from micro- and macro-telluric absorption in the Earth's atmosphere
- Evaluate the unique advantages that a space-based platform provides to enable the identification and mitigation of stellar activity for multi-planet signal recovery

**Mission Concept Overview:**
The primary science goals of EarthFinder are the radial velocity detection, mass measurement, and orbit characterization of Earth-mass planets in Habitable Zone orbits around the nearest FGKM stars. These goals correspond to a radial velocity precision of 1 cm/s on time-scales of several years, given the 9 cm/s reflex motion velocity semi-amplitude of the Sun in response to Earth and a 10% mass determination precision (Figure 1). Many ancillary science cases would be possible, including direct exoplanet spectroscopy, stellar dynamos and asteroseismology, fundamental atomic transitions in the Sun and other stars, following the water in the local Universe obscured by telluric water, brown dwarf atmospheres, the study of diffuse interstellar bands, the direct detection of the acceleration of the Universe, and more since a spectrometer with such a high resolution has never flown in space before.

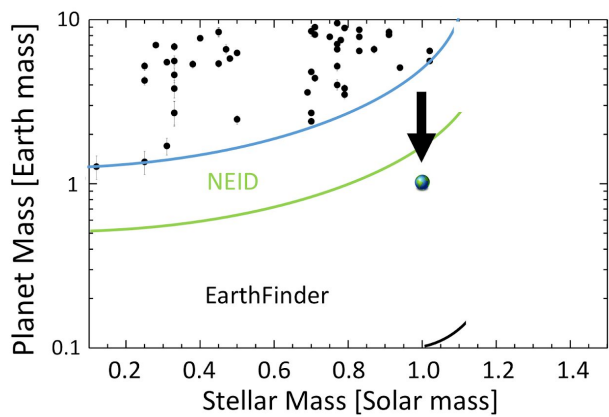

*Figure 1* PRV-discovered exoplanets less than 10 $M_{Earth}$ as a function of stellar mass and planet mass modulo the unknown inclination. Black circles are data from the NASA Exoplanet Archive. The blue-green orb corresponds to the Earth. The blue curve corresponds to the approximate current detection limit of the PRV method, the green curve corresponds to the NEID spectrometer (or similarly, EXPRES, or ESPRESSO), and the black curve corresponds to EarthFinder.

The nominal spacecraft design is based upon the Kepler spacecraft by Ball Aerospace, with a 1.4-m primary, with the starlight coupled into single-mode fibers illuminating three high-resolution, compact and diffraction-limited spectrometer "arms", one covering the near-UV (200-380 nm), visible (380-900 nm) and near-infrared (NIR; 900-2500 nm) respectively with a spectral resolution of greater than 150,000 in the visible and near-infrared arms (Figure 2). A small Solar telescope near the solar panels would also be included to obtain Solar spectra.

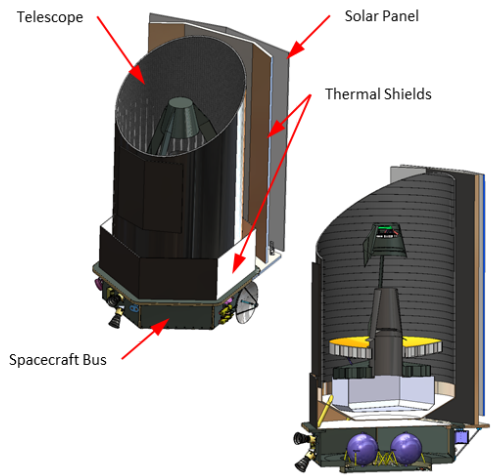

*Figure 2: Spacecraft concept views. The bay between primary mirror and bus is for instrumentation (~0.5m$^3$).*

The telescope would be baffled to minimize the Sun avoidance angle, and with thermal shielding on the spacecraft bus to minimize the anti-Sun avoidance angle and to maximize the instantaneous field of regard. The spacecraft would be placed in an Earth-trailing or Lagrange orbit for minimal velocity changes with respect to the Solar System barycenter during science exposures, with a primary science mission duration of 5 years. Many of the currently known PRV error terms can be eliminated or mitigated by going to space (Figure 3).

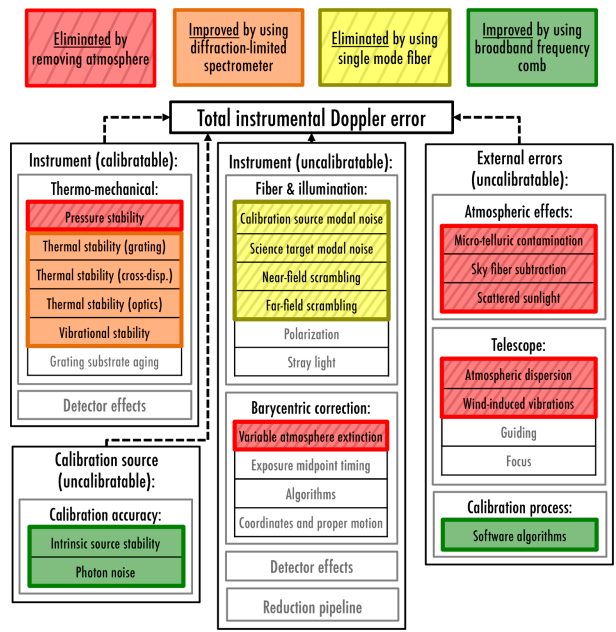

*Figure 3: Highlighted terms represent errors that are either significantly reduced, or entirely eliminated, by (1) removing the atmosphere (red), (2) using a compact, diffraction-limited spectrometer (orange), (3) delivering light to the instrument using a single-mode optical fiber (yellow), or (4) using a broadband optical frequency comb calibration source.* **Removing or significantly reducing these instrumental error sources opens an entirely new discovery space.**

**Survey & Cadence**

EarthFinder would survey the nearest 25-50 FGKM dwarf stars with a cadence unimpeded by diurnal and seasonal cycles and aliases (no Rayleigh-scattering atmosphere), and without declination limits and right ascension target bias (shorter nights lead to fewer epochs per target during the summer). The one-day cadence aliasing from ground-based facilities draws significant power away from the exoplanet signals (Figure 4). The seasonal gaps are very important and under-appreciated for limiting sensitivity to the habitable zone exoplanet orbital periods of 0.5-2 years. The lack of "daytime" and weather offers a gain in available integration time of a factor more than three to partially offset the smaller aperture.

**Stellar Activity:**

Dynamics on the surface of the star introduce apparent false velocity changes, and the identification and modeling of stellar activity from cool spots, plages, granulation, and pulsations is an area of active research (Dumusque et al. 2017). How well we can correct for stellar activity below 1 m/s is not yet known because we have not had the instruments with the necessary stability and characteristics to find out until this year (e.g. ESPRESSO, EXPRES, and NEID, Fischer et al. 2016). A variety of techniques have been proposed to mitigate stellar activity - including high cadence observations (e.g. MINERVA, and see Barnes et al. 2017, Rajpaul et al. 2017), simultaneous visible and NIR arms (e.g. CARMENES, see the RV-color index in Tal-Or et al. 2018 and StarSIM simulations in Herrero et al. 2016), simultaneous space-based photometry (e.g. RVxK2, and Oshagh et al. 2017), extreme spectral resolution (R>150k, e.g. iLocater; see Figure 3 in Jurgenson et al. 2016), line-by-line analysis (Dumusque et al. priv. comm.), traditional activity indicators (CCF FWHM, CCF Bisector, CaII R'HK, Hα), and potentially polarization (e.g. SPIRou). Some of these techniques such as high cadence observations and spectral grasp have shown initial promise at partially mitigating stellar activity, while others (e.g. extreme resolution) remain to be tested. EarthFinder in space is the only platform that uniquely enables all of these approaches for mitigating stellar activity, particularly the cadence as previously discussed and telluric-free red/NIR velocities.

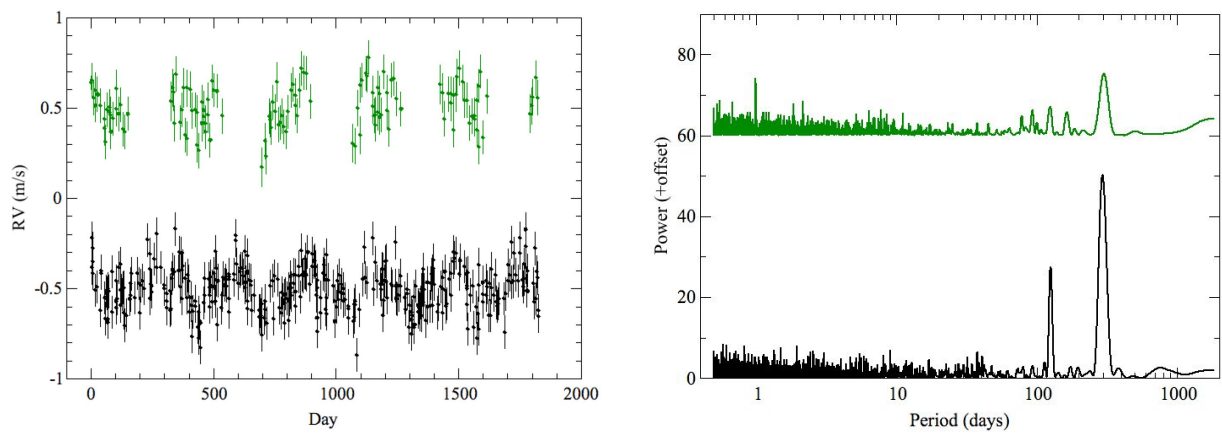

*Figure 4:* Left: Green: Simulated cadence for the target with the median number of observations from a 42-target 5 year survey with a "super-NEID" on the WIYN with 3 cm/s photon+instrument noise, and stellar activity "corrected" to white noise at 9 cm/s, accounting for sunrise and sunset, airmass, weather loss, etc. ($N_{obs}$=137). Black: The same target as seen from EarthFinder in the continuous viewing zone, simulated with a uniform random cadence ($N_{obs}$=396). Two terrestrial mass planets are injected, one at ~300 days (K=9 cm/s) and one at ~120 days (K=7 cm/s). Right: Periodograms for both ground (green) and space (black) cadences. The space-based cadence has a false alarm probability $10^{14}$ times lower for the longer-period planet, and $10^8$ times lower for the shorter period planet, the latter of which is essentially not detected from the ground. Several diurnal and seasonal period aliases are present in the ground-based cadence. While this appears to be an unfair comparison in terms of the number of observations, this simulated target is only visible from the ground 14.4% of the time, after accounting for daytime, seasonal visibility and weather losses. Thus the number of observations per visible unit of time is actually 76% higher from the ground for this particular example.

We are in the process of simulating our ability to recover injected stellar activity signals into RV time-series data from the ground and space using the StarSIM 2.0 code and Gaussian processes (Herrero et al. 2016). For EarthFinder, the RV color from activity is readily detected between the visible and NIR arms (Figure 5).

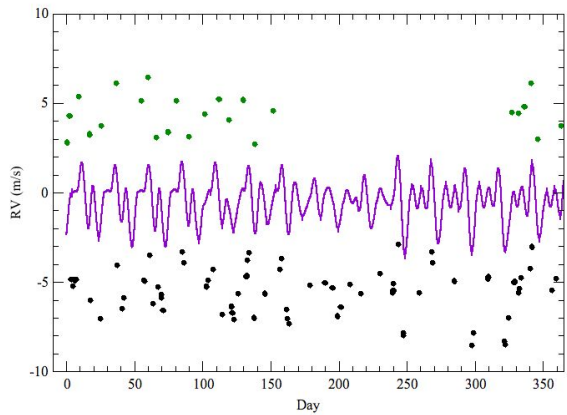

*Figure 5:* Purple: One year of simulated RV color ($RV_{VIS}$-$RV_{NIR}$) time-series for an active Sun. Green: The simulated ground-based cadence + 5 m/s offset assuming perfect telluric correction; Black: The simulated space-based cadence, -5 m/s offset.

**Telluric Absorption:**

Telluric absorption poses a serious challenge to PRVs. It is a known bottleneck for achieving higher RV precision (<3 m/s) in the NIR (Bean et al. 2010; Reiners et al. 2017). Moreover, even the "micro-telluric" lines (depths <2% and mostly <1%) in the visible can contribute to the RV error budget at the 20-50 cm/s level (Cunha et al. 2014; Artigau et al. 2014). A recent work by Sithajan et al. (2016) concluded that, even if all of the telluric lines are modeled and subtracted to the 1% level (which is extremely challenging; e.g., Seifahrt et al. 2010, Gullikson et al. 2014, Smette et al. 2015), the residuals would still cause 0.4-1.5 m/s RV errors in the NIR for M and K dwarfs. This is a large term in the PRV error budget which can be eliminated by moving to space.

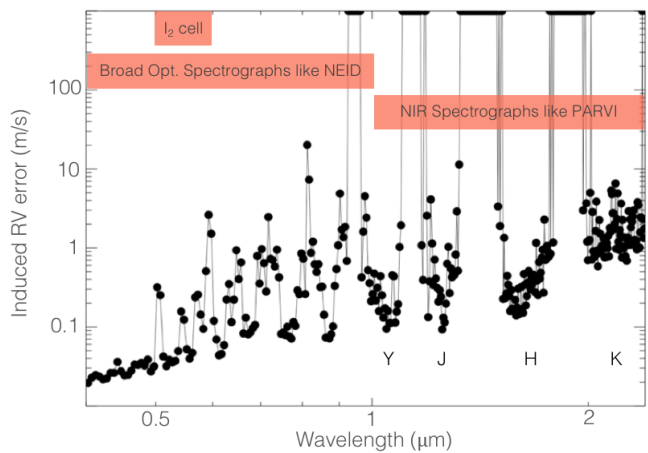

*Figure 6:* RV error induced by residuals in correcting telluric absorption lines in each 5 nm spectral chunk (simulation from Cullen Blake). Telluric absorption lines are corrected by directly dividing them out in the simulated observed spectra, with a known, constant LSF.

The goal of our study is to explore the best strategies for mitigating telluric contamination and quantify their effectiveness. We will also investigate the RV errors induced by the time variability of tellurics, especially considering that cm/s photon noise requires long exposures.

**One Path Forward**

In addition to the lessons we will learn from the next generation of visible and NIR PRV instruments, (Wright & Robinson 2017, Plavchan et al., 2015), we need experimental data to prepare for the next generation of spectrometers that will reach 1 cm/s. We outline a partial list of possible experiments in the near-term to refine the EarthFinder mission science objectives and feasibility, and the next generation of PRV spectrometers in general:

● A diffraction-limited test-bed echelle spectrometer and Fourier Transform Spectrometer, both fed by a Solar telescope analogous to the one for HARPS-N, with the echelle spectrometer continually modified and upgraded from the current best practices of 20 cm/s to 1 cm/s to understand the known and unknown error terms in the error budget.

- A high-resolution, stabilized red/NIR spectrometer for SOFIA to evaluate the effectiveness of (mostly) removing the Earth's atmosphere.
- A balloon-based high-resolution, diffraction-limited spectrograph with disk-integrated Sunlight coupled via an integrating sphere or via a small Solar telescope (<10 cm).
- A SmallSat ($30M; e.g., VELOS SmallSat concept white paper; PI: Beichman) carrying a diffraction-limited spectrograph with disk-integrated Sunlight coupled via an integrating sphere.

We need to demonstrate that we can correct the activity for the Sun at the <10 cm/s level before we can attempt to do so for other stars. Existing heliophysics satellites are not adequate - SDO/HMI photospheric velocities are derived from a single iron line with 7 m/s precision.

It is essential to couple a robust technology development plan with investments in data analysis techniques, simulations, and partnerships with heliophysics. PRVs are in transition from PI-based science to "Big Science." To ensure the future utility of the PRV method for exoplanet science, a centralized and coordinated effort is needed, and a fundamentally different funding model than exists today. A single "dream" RV machine on a single dedicated ground-based 8-m class telescope could potentially cost $50-$100M, and a global network of dedicated RV facilities could approach the cost of a probe mission anyway, potentially on the order of $0.5B. If this effort is successful, and we are able to discover the HZ Exo-Earths around nearby FGKM stars with PRVs, the cost could potentially pay for itself in the architectural savings realized from a future flagship direct-imaging mission that does not need to conduct a blind survey with multiple revisits, and instead knows which stars to look at, when the planets are at quadrature, and which planet is which. Either way, 1 cm/s PRVs will be needed for the orbit and 10% mass characterization of directly-imaged Exo-Earths.